\documentclass[aps,pre,showpacs,reprint,groupedaddress]{revtex4-2}

\usepackage{amsmath}
\usepackage{lmodern}
\usepackage{graphicx}
\usepackage{amssymb} 
\usepackage{physics}
\usepackage{color}
\usepackage[colorlinks,citecolor=blue,breaklinks=true]{hyperref}

\DeclareUnicodeCharacter{03B2}{\ensuremath{\beta}}
\begin{document}
%opening
\title{Suppression of synchronous spiking in two interacting populations of excitatory and inhibitory quadratic integrate-and-fire neurons} 

\author{Kestutis~Pyragas, Augustinas~P.~Fedaravi\v{c}ius, and Tatjana~Pyragien\.{e}}
\affiliation{Center for Physical Sciences and Technology, LT-10257 Vilnius, Lithuania}
%\author{Kestutis~Pyragas}
% \email{kestutis.pyragas@ftmc.lt}
 %\homepage[]{http://pyragas.pfi.lt}
%\thanks{}
% \affiliation{Center for Physical Sciences and Technology, LT-10257 Vilnius, Lithuania}

%\author{Augustinas~P.~Fedaravi\v{c}ius}
% \affiliation{Center for Physical Sciences and Technology, LT-10257 Vilnius, Lithuania}
 
%\author{Tatjana~Pyragien\.{e}}
% \affiliation{Center for Physical Sciences and Technology, LT-10257 Vilnius, Lithuania}

\begin{abstract}

Collective oscillations and their suppression by external stimulation are analyzed  in a large-scale neural network consisting of two interacting populations of excitatory and inhibitory quadratic integrate-and-fire neurons. In the limit of an infinite number of neurons, the microscopic model of this network can be reduced to an exact low-dimensional system of mean-field equations. Bifurcation analysis of these equations reveals three different dynamic modes in a free network: a stable resting state, a stable limit cycle, and bistability with a coexisting resting state and a limit cycle. We show that in the limit cycle mode, high-frequency stimulation of an inhibitory population can stabilize an unstable resting state and effectively suppress collective oscillations. We also show that in the bistable mode, the dynamics of the network can be switched from a stable limit cycle to a stable resting state by applying an inhibitory pulse to the excitatory population. The results obtained from the mean-field equations are confirmed by numerical simulation of the microscopic model.
\end{abstract}


\maketitle

\section{Introduction}
\label{sec:introduction}

Synchronization processes in large populations of interacting dynamical units are the focus of intense research in physical, technological and biological  systems~\cite{Arenas2008}, such as smart grids~\cite{Dorfler2013}, Josephson junction arrays~\cite{Galin2018}, coupled mechanical devices~\cite{Zhang2012}, optical networks~\cite{Hagerstrom2012}, and neural networks~\cite{Guevara2017}. In neural networks, synchronization can play a dual role. Under normal conditions, synchronization is responsible for cognition and learning~\cite{Singer1999,Fell2011}, while excessive synchronization can cause abnormal brain rhythms associated with neurological diseases such as Parkinson's disease~\cite{Hammond2007}, epilepsy~\cite{Jiruska2013,Gerster2020}, tinnitus~\cite{Tass2012tin}, and others. Various open loop and closed loop control algorithms have been developed to suppress unwanted synchronized network oscillations, e.g., coordinated reset stimulation~\cite{tass2003,Popovych2014}, time-delayed feedback control~\cite{Rosenblum2004,Popovych2005,Popovych2017}, separate stimulation-registration setup~\cite{Pyragas2007}, act-and-wait algorithm~\cite{Ratas2014,Ratas2016ND}, optimal open loop desynchronization~\cite{Wilson2020}, and many others. 

A therapeutic procedure clinically approved for the treatment of Parkinson's disease, essential tremor and dystonia  is a high-frequency (HF) deep brain stimulation (DBS) \cite{Benabid1991,kring07}. The mechanism of action of DBS is still poorly understood~\cite{Lozano2004,Lozano2019}. Clinical observations show that the effects of lesions and DBS of the same target area are similar~\cite{Limousin1995}. This suggests that HF stimulation suppresses neuronal activity in the target area. The hypothesis of the local inhibition is also supported by some experiments in animals~\cite{Benazzouz2000,Tai2003} and in humans~\cite{Dostrovsky2000,Filali2004}. In this context, the effect of HF stimulation can be explained in terms of stabilization of neuron's resting state~\cite{Pyragas2013}. However, there is no clear theoretical understanding of how HF stimulation affects synchronization processes in neural networks.

Recent advances in dynamical systems theory have allowed us to better understand the effects of synchronization in large-scale oscillatory networks. A major breakthrough in these studies was achieved by Ott and Antonsen~\cite{Ott2008}, who showed that the microscopic model equations of globally coupled heterogeneous phase oscillators (Kuramoto model) can be reduced to a low-dimensional system of ordinary differential equations that accurately describe the macroscopic evolution of the system in the infinite-size (thermodynamic) limit. Later this approach was extended to a particular class of heterogeneous neural networks composed of all-to-all pulse-coupled quadratic integrate-and-fire (QIF) neurons~\cite{Montbrio2015}, which are the normal form of class I neurons~\cite{izhi07}. In thermodynamic limit, a low-dimensional system of mean-field equations was derived for biophysically relevant macroscopic quantities: the firing rate and the mean membrane potential. The approach has been further developed in recent publications to analyze the occurrence of synchronized macroscopic oscillations in networks of QIF neurons  with a realistic synaptic coupling~\cite{Ratas2016}, in the presence of a delay in couplings~\cite{Pazo2016,Devalle2017,Ratas2018}, in the presence of noise~\cite{Ratas2019}, in the presence of electrical coupling~\cite{Montbrio2020} and in the case of two interacting populations~\cite{Ratas2017,Segneri2020}. 

In this paper, we demonstrate that mean-field equations are useful not only for understanding the occurrence of collective oscillations in large-scale neural networks, but also for understanding the effect of stimulation on synchronization processes. As an example, we consider a network of two interacting populations of excitatory and inhibitory QIF neurons. We show that HF stimulation of the inhibitory population is very effective in suppressing the collective synchronous spiking in both populations. The suppression mechanism is explained by the stabilization of the unstable incoherent state of the network. We also explain the oscillation suppression effect caused by an inhibitory pulse applied to the excitatory population.

The rest of the paper is organized as follows. In Sec.~\ref{sec:model} we describe a microscopic model of two interacting populations of QIF neurons and present the reduced mean-field equations for this model. Section~\ref{sec:free_dynamics} is devoted to bifurcation analysis of the mean-field equations without stimulation. The effects of HF stimulation of the inhibitory population, as well as the inhibitory pulse applied to the excitatory population, are discussed in Sec.~\ref{sec:suppression}.
%
%In Sec.~\ref{sec:suppression} we discuss the effects of stimulation. To explain the suppression of synchronous spiking by HF stimulation, we use mean-field equations averaged over the stimulation period. We also consider the effect of an inhibitory pulse applied to the excitatory population. 
%
In Sec.~\ref{sec:microscopic_model}, we present the results of numerical simulations of the microscopic model and compare them with the results obtained from the mean-field equations. The conclusions and discussion  are presented in Sec.~\ref{sec:conclusions}.

\section{The model}
\label{sec:model}

\subsection{Microscopic description}
\label{microscopic}
We consider a heterogeneous network of two interacting populations of excitatory and inhibitory quadratic integrate-and-fire neurons, which are the canonical representatives for  class I neurons near the spiking threshold~\cite{izhi07}.  The microscopic state of the network is determined by the set of $2N$ neurons' membrane potentials  $\{V_j^{(E,I)}\}_{j=1,\ldots,N}$, which satisfy the following system of $2N$ ordinary differential equations \cite{ermentrout10}:
\begin{eqnarray}
\tau\dot{V}_{j}^{(E,I)} &=&  ({V}_{j}^{(E,I)})^{2}+\eta_j^{(E,I)}+\mathcal{I}_{j}^{(E,I)},\nonumber\\
& & \;\; \text{if} \;\; {V}_{j}^{(E,I)}\ge V_{p} \;\; \text{then} \;\; {V}_{j}^{(E,I)} \leftarrow V_{r}. \label{model}
\end{eqnarray}
Here, $\tau$ is the membrane time constant and $V_j^{(E,I)}$ is the membrane potential of neuron $j$ in either the excitatory $(E)$ or the inhibitory $(I)$ population. For simplicity, we set the number of neurons $N$ and the time constant $\tau$ the same for both populations. The heterogeneous parameter of excitability $\eta_j^{(E,I)}$  is a current that specifies the behavior of each isolated neuron and the term $\mathcal{I}_{j}^{(E,I)}$ defines the synaptic coupling between neurons as well as external stimulation. The isolated neurons ($\mathcal{I}_{j}^{(E,I)}=0$) with the negative value of the parameter $\eta_j^{(E,I)}<0$ are at rest, while the neurons with the positive value of the parameter $\eta_j^{(E,I)}>0$ generate instantaneous spikes, which are approximated by the Dirac delta function. The spikes are emitted at the moments when the membrane potential $V_j^{(E,I)}$ reaches a peak value $V_{p}$. Immediately after the spike emission the membrane potential is reset to a value $V_{r}$. Thereafter, we assume $V_{p}= -V_{r} \to \infty$. With this assumption, a QIF neuron can be transformed into a theta neuron. This assumption is also crucial for the analytical treatment of the Eqs.~\eqref{model}  in an infinite  size limit $N \to \infty$ \cite{Montbrio2015}. The values of the heterogeneous parameter $\eta_j^{(E,I)}$ for both populations are independently taken  from the Lorentzian distributions:
\begin{equation}
g_{E,I}(\eta)=\frac{1}{\pi} \frac{\Delta_{E,I}}{(\eta-\bar{\eta}_{E,I})^2+\Delta_{E,I}^2}, \label{Lor}
\end{equation}
where $\Delta_{E,I}$ and $\bar{\eta}_{E,I}$ are respectively the width and the center of the distribution for the excitatory (E) and inhibitory (I) populations.
\begin{figure}
\centering
	\includegraphics{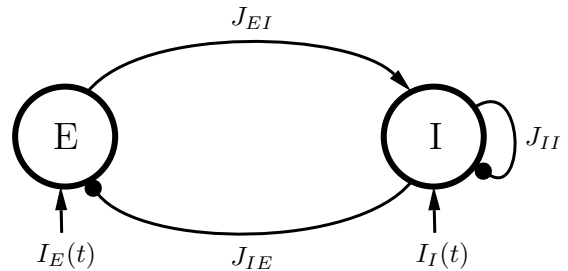}
\caption{\label{EI_network} Symbolic depiction of a network of two neural populations.  The large circles labeled ``E'' and ``I'' represent populations of excitatory  and inhibitory  neurons, respectively. The curve ending in an arrow shows excitatory  coupling between populations E and I. $J_{EI}$ is the coupling strength. Curves ending in solid circles indicate inhibitory couplings between populations I and E, as well as within population I. $J_{IE}$ and $J_{II}$ are the corresponding  coupling strengths. The vertical arrows labeled $I_E(t)$ and $I_I(t)$ show the external stimulation currents applied to populations E and I. 
}
\end{figure}
%

%The isolated [$\mathcal{I}_{j}^{(E,I)}=0$] QIF neuron is the canonical model for the class I neurons near the spiking threshold~\cite{izhi07}. Spiking instability in such neurons is manifested through bifurcation of the saddle node on the invariant curve (SNIC). The system following this scenario exhibits excitability before the bifurcation. For the QIF neuron, this scenario is provided by the bifurcation parameter $\eta_j^{(E,I)}$. For $\eta_j^{(E,I)}<0$, the neuron is in the excitable mode and for $\eta_j^{(E,I)}>0$ it is in the spiking mode. 

Finally, we discuss the last term $\mathcal{I}_{j}^{(E,I)}$ in Eqs.~\eqref{model}, which describes synaptic coupling and an external stimulation. For the excitatory and inhibitory populations this term respectively is
\begin{subequations}
\label{currents}
\begin{eqnarray}
\mathcal{I}_{j}^{(E)} & = & -J_{IE}S_I(t)+I_E(t),\label{current_e}\\
\mathcal{I}_{j}^{(I)} & = & J_{EI}S_E(t)-J_{II}S_I(t)+I_I(t).\label{current_i}
\end{eqnarray}
\end{subequations}
Here, $S_E(t)$ and $S_I(t)$ determine the mean synaptic activation of E and I populations:
\begin{equation}
S_{E,I}(t) = \frac{\tau}{N}\sum_{j=1}^{N}\, \sum_{k\setminus (t_{j}^{k})_{E,I}<t}\delta(t-(t_{j}^{k})_{E,I}), \label{SEI}
\end{equation}
where $(t_{j}^{k})_{E,I}$ is the time of the $k$th spike of the $j$th neuron in either E or I population and $\delta(t)$ is the Dirac delta function. The positive parameters $J_{EI}$, $J_{IE}$ and $J_{II}$ define synaptic weights. The current $-J_{IE}S_I(t)$ inhibits E neurons due to synaptic activity of I population, while the current $J_{EI}S_E(t)$ excites I neurons due to synaptic activity of E population.  The term $-J_{II}S_I(t)$ determines recurrent inhibition of neurons within I population. For simplicity, we do not consider recurrent excitation within the E population, since it is not essential for the emergence of collective oscillations. The currents $I_E(t)$ and $I_I(t)$ represent external homogeneous  stimulation of the excitatory and the inhibitory populations, respectively. Below we will consider stimulation protocols when either only inhibitory ($I_E(t)=0$, $I_I(t)\neq 0$) or only excitatory ($I_E(t)\neq 0$, $I_I(t)=0$) population is stimulated.

Note that the dynamics of a single population of QIF neurons interacting via instantaneous Dirac delta pulses was studied in detail in Ref.~\cite{Montbrio2015} and macroscopic limit cycle oscillations were not found in such a model. Macroscopic synchronized oscillations can occur in a single population when there is a delay in couplings~\cite{Pazo2016,Devalle2017,Ratas2018} or when the finite width of synaptic pulses is taken into account~\cite{Ratas2016}. However, two coupled populations of excitatory and inhibitory QIF neurons can generate macroscopic oscillations even when the interaction is provided by instantaneous Dirac delta pulses~\cite{Segneri2020}, and therefore we restrict our consideration to the simpler case of instantaneous interaction, as in the original paper~\cite{Montbrio2015}.

The network architecture shown in Fig.~\ref{EI_network}, mimics the architecture of the neural network of the subthalamic nucleus (STN) and the external segment of the globus pallidus (GPe), which is often used to model Parkinson's disease (cf.,e.g., Ref.~\cite{Terman2002}). STN is a network of excitatory neurons (in our case E population), and GPe consists of  inhibitory neurons (in our case I population).

\subsection{Macroscopic description: Low-dimensional mean-field equations in the limit $N\to \infty$ }
\label{sec:_tetod_limit}

The advantage of the network  model Eqs.~\eqref{model} is that it allows one to derive precise low-dimensional mean-field equations in the thermodynamic limit of an infinite number of neurons, $N \to \infty$. We characterize the macroscopic dynamics of the network by four biophysically relevant quantities
\begin{equation}
v_{E,I}=\frac{1}{N}\sum_{j=1}^N V_j^{(E,I)}, \quad r_{E,I}=  \tau\frac{M^{(E,I)}(\Delta t)}{N \Delta t},  \label{rv_dif}
\end{equation}
which represent the  mean membrane potentials of the excitatory (E) and the inhibitory (I) populations and the dimensionless  firing rates of E and I populations (the dimensional firing rates are $r_{E,I}/\tau$), respectively.  Here, $M^{(E,I)}(\Delta t)$ is the number of spikes emitted in a small time window $\Delta t$ in $E$ and $I$ populations. In the limit $N \to \infty$, the quantities $r_{E,I}(t)$ and $v_{E,I}(t)$ satisfy the exact system of four ordinary differential equations~\cite{Montbrio2015}:
\begin{subequations}
\label{eq_rvEI}
\begin{eqnarray}
\tau\dot{r}_E & = & \Delta_E/\pi+ 2r_Ev_E, \label{eq_rE}\\
\tau\dot{v}_E & = & \bar{\eta}_E +v_E^2-\pi^2 r_E^2-J_{IE}r_I+I_E(t),\label{eq_vE}\\
\tau\dot{r}_I & = & \Delta_I/\pi+ 2r_Iv_I, \label{eq_rI}\\
\tau\dot{v}_I & = & \bar{\eta}_I +v_I^2-\pi^2 r_I^2+J_{EI}r_E-J_{II}r_I+I_I(t). \label{eq_vI}
\end{eqnarray}
\end{subequations}
These low-dimensional mean-field equations greatly simplify the analysis of different network dynamics modes depending on system parameters, as well as the effect of stimulating current on network dynamics.

\section{Network dynamics without stimulation}
\label{sec:free_dynamics}

\begin{figure}
\centering
	\includegraphics{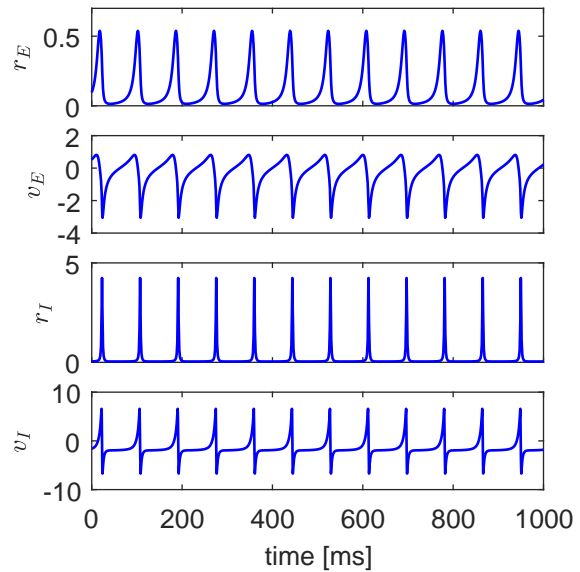}
\caption{\label{free_netw_dyn} Dynamics of network macroscopic  variables $r_{E}(t)$, $v_{E}(t)$, $r_{I}(t)$ and $v_{I}(t)$ obtained by solving the Eqs.~\eqref{eq_rvEI} without stimulation, $I_E(t)=I_I(t)=0$. The values of the parameters are: $\Delta_E=0.05$, $\bar{\eta}_E=0.5$, $\Delta_I=0.5$, $\bar{\eta}_I=-4$, $J_{EI}=20$, $J_{IE}=5$, $J_{II}=0.5$ and $\tau=14$~ms. 
}
\end{figure}

First, we analyze the dynamics of the network without stimulation, $I_E(t)=I_I(t)=0$. An unperturbed network exhibits synchronized oscillations over a wide range of parameters. An example of network oscillations obtained by solving the Eqs.~\eqref{eq_rvEI} for the set of parameters  $\Delta_E=0.05$, $\bar{\eta}_E=0.5$, $\Delta_I=0.5$, $\bar{\eta}_I=-4$, $J_{EI}=20$, $J_{IE}=5$ and $J_{II}=0.5$  is shown in Fig.~\ref{free_netw_dyn}. To get oscillations on a realistic time scale, we choose $\tau=14$~ms, which corresponds to the membrane time constant of GPe neurons~\cite{Kita1991}. The oscillation period in the figure is $T_0 \approx 87$ ms. 

Areas in the parameter space where network oscillations occur can be estimated using linear stability analysis of the Eqs.~\eqref{eq_rvEI}. 
Equating the right-hand side (RHS) of Eqs.~\eqref{eq_rvEI} to zero, we can find the fixed points $(r_E^*, v_E^*, r_I^*, v_I^*)$ in the four dimensional phase space of the system. The problem leads to the solution of the 16th order polynomial equation 
\begin{equation}
a_1 [P(x)]^4-(1+x)x^4 [P(x)]^2 -a_2 x^8+a_3x^2 [P(x)]^3=0  \label{fp_eq}
\end{equation}
with respect to $x$, where
\begin{equation}
P(x)= a_5+x^2-a_4 x^4  \label{polin}
\end{equation}
is a fourth order polynomial and the parameters $a_j$ are:
\begin{align*}
a_1&=\frac{1}{\bar{\eta}_I} \left(\frac{\pi \bar{\eta}_E}{J_{IE}}\right)^2, &  
a_2&=\frac{1}{\bar{\eta}_I} \left(\frac{J_{IE} \Delta_I}{2\pi \bar{\eta}_E}\right)^2, &              
a_3&=\frac{\bar{\eta}_E J_{II}}{\bar{\eta}_I J_{IE}},\\
a_4&= \frac{1}{\bar{\eta}_E} \left(\frac{\pi \bar{\eta}_I}{J_{EI}}\right)^2, &    
a_5&=\frac{1}{\bar{\eta}_E} \left(\frac{J_{EI} \Delta_E}{2\pi \bar{\eta}_I}\right)^2.  
\end{align*}  
The solutions of the polynomial Eq.~\eqref{fp_eq} are related to the coordinates of fixed points as follows:
\begin{align*}
r_E^* &=\frac{\bar{\eta}_I}{J_{EI}} x, &  
v_E^* &=-\frac{J_{EI}\Delta_E}{2\pi\bar{\eta}_I} \frac{1}{x}, \\
r_I^* &=\frac{\bar{\eta}_E}{J_{IE}} \frac{P(x)}{x^2}, &  
v_I^* &=-\frac{J_{IE}\Delta_I}{2\pi\bar{\eta}_E} \frac{x^2}{P(x)}. 
\end{align*}  
Numerical analysis of  Eq.~\eqref{fp_eq} shows that only one of its real-valued roots satisfies the requirement of non-negativity of spiking rates $r_E^*$ and $r_I^*$, that is, the system has a single fixed point in the physically relevant region $r_E \geq 0$ and $r_I \geq 0$ of the phase space.
The stability of this fixed point is determined by the eigenvalues $\lambda$ of the characteristic equation 
\begin{equation}
\det(\mathcal{J}-I\lambda)= 0,  \label{charact_eq}
\end{equation}
where 
\begin{equation}
\mathcal{J}=\frac{1}{\tau}
\begin{pmatrix}
   2v_E^* & 2r_E^* & 0 & 0 & \\ 
 -2\pi^2r_E^* &  2v_E^* & -J_{IE} & 0 \\
  0 & 0 & 2v_I^* & 2r_I^*\\ 
  J_{EI} & 0 & -(2\pi^2r_I^*+J_{II}) & 2v_I^*
\end{pmatrix}  \label{jacobian}
\end{equation}
is the Jacobi matrix of the system Eqs.~\eqref{eq_rvEI} and $I$ is the identity matrix. The fixed point is stable if the real parts of all eigenvalues $\lambda$ are negative. With such parameter values, the network is at rest. At the microscopic level, neurons in this state exhibit incoherent behavior. If the real part of at least one of the eigenvalues $\lambda$ is positive, than the state of rest becomes unstable. Numerical analysis shows that in this case neurons behave coherently and periodic limit cycle oscillations appear in the network. Thus, solving the Eqs.~\eqref{fp_eq} and \eqref{charact_eq} gives us a simple sufficient condition for identifying areas in the parameter space where the network is in oscillatory mode. However, network oscillations can also occur in parameter areas where the resting state is stable. In such areas, the network demonstrates bistability. Along with a stable resting state, the network has a stable limit cycle. 

\begin{figure}
\centering
	\includegraphics{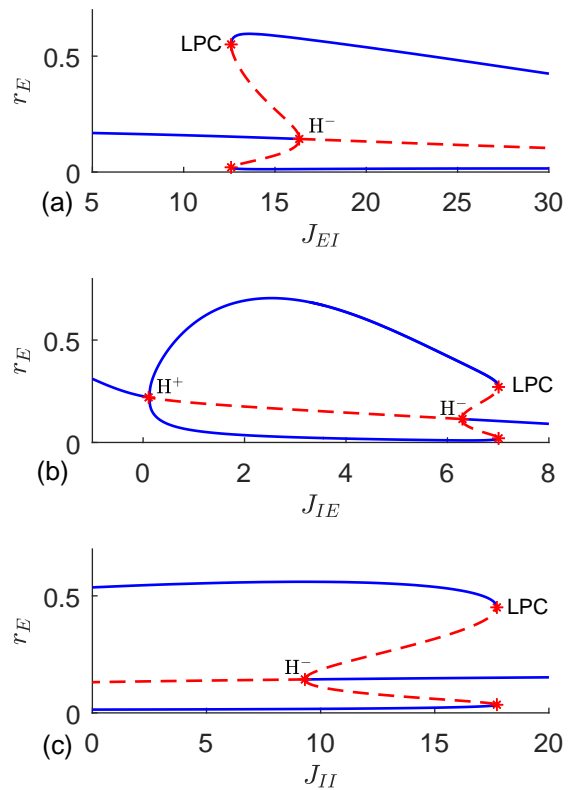}
\caption{\label{bif_onepar3} One-parameter bifurcation diagrams showing the evolution of the firing rate $r_E$ depending on the coupling strengths (a) $J_{EI}$, (b) $J_{IE}$ and (c) $J_{II}$. The rest of the parameters is fixed in the same way as in Fig.~\ref{free_netw_dyn}. The solid blue curves show the stable fixed point and the maximum and minimum of the stable limit cycle. 
The dashed red curves correspond to the unstable fixed point and the maximum and minimum of the unstable limit cycle. The red asterisks marked with the letters LPC, H$^+$  and H$^-$ denote the limit point of cycles bifurcation, supercritical Hopf bifurcation and subcritical Hopf bifurcation, respectively. 
}
\end{figure}

The bistability regions are clearly visible in the one-parameter bifurcation diagrams shown in Fig.~\ref{bif_onepar3}. These and other bifurcation diagrams presented in this paper were built using the MatCont package~\cite{matcont}. Figure ~\ref{bif_onepar3}(a) shows change in  the firing rate $r_E$ of the excitatory population depending on the coupling strength $J_{EI}$. There are two bifurcation values of the coupling strength, at which the dynamics of the network changes qualitatively. At $J_{EI}\approx 12.6$, there is a limit point of cycles (LPC) bifurcation when two limit cycles, stable and unstable, collide and annihilate each other. At $J_{EI}\approx 16.35$, a subcritical Hopf (H$^-$) bifurcation occurs when an unstable limit cycle is absorbed by a stable spiral equilibrium. As $J_{EI}$ increases, different dynamic modes are observed. For small values of the coupling strength $J_{EI}< 12.6$, the resting state is the only attractor. In the interval $12.6<J_{EI}< 16.35$ between the LPC and H$^-$ bifurcations, there are two attractors: a state of rest and a limit cycle. 
Finally, for $J_{EI}> 16.35$, the only attractor is the limit cycle. Figure~\ref{bif_onepar3}(b) shows the evolution of network dynamics with a change in the coupling strength $J_{IE}$. As this parameter increases from zero, the oscillations manifest themselves through the supercritical Hopf (H$^+$) bifurcation at $J_{IE}\approx 0.13$. In the interval $0.13<J_{IE}<6.28$ between the supercritical and subcritical Hopf bifurcations, the only attractor is the limit cycle. In the interval  $6.28<J_{IE}<7$ between the bifurcations H$^-$ and LPC there is  bistability, and for $J_{IE}>7$ the rest state is the only attractor. Figure~\ref{bif_onepar3}(c) shows that oscillations in the network occur with zero interaction within the inhibitory population, $J_{II}=0$. As $J_ {II}$ increases, the oscillations persist until the LPC bifurcation, $J_ {II}=17.72$. At $J_{II}>17.72$, the oscillations disappear, and the only attractor is the state of rest. The bistability is in the interval $9.3<J_{II}<17.72$ between the H$^-$ and LPC bifurcations.  

\begin{figure}
\centering
	\includegraphics{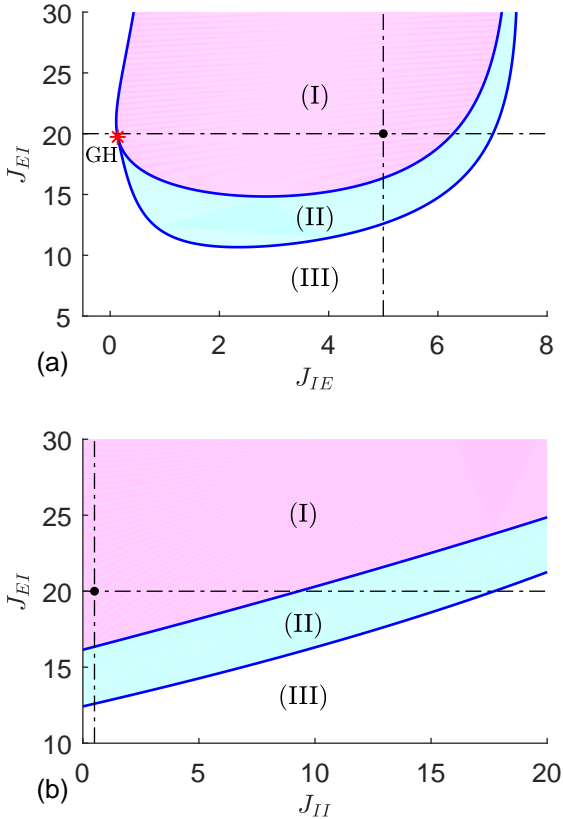}
\caption{\label{bif_twopar2} Two-parameter bifurcation diagrams in the planes of parameters (a) ($J_{IE}, J_{EI}$) and (b) ($J_{II}, J_{EI}$). Other parameters are the same as in Fig.~\ref{free_netw_dyn}. The areas marked with Roman numerals correspond to: (I) - the only stable limit cycle, (II) - bistability with a stable limit cycle and a stable resting state, and (III) - the only stable resting state. The red asterisk marked with letters GH denotes the generalized Hopf bifurcation point. The horizontal dash-dotted lines $J_{EI}=20$ in (a) and (b) correspond to the one-parameter bifurcation diagrams shown in Figs.~\ref{bif_onepar3}(b) and \ref{bif_onepar3}(c), respectively. The vertical dash-dotted lines $J_{IE}=5$ in (a) and $J_{II}=0.5$ in (b) correspond to the one-parameter bifurcation diagram shown in Fig.~\ref{bif_onepar3}(a). The intersection of the horizontal and vertical lines represents the parameter values used in Fig.~\ref{free_netw_dyn}.
}
\end{figure}

In Figs.~\ref{bif_twopar2}(a) and \ref{bif_twopar2}(b), we present two-parameter bifurcation diagrams in the parameter planes ($J_{IE}, J_{EI}$) and  ($J_{II}, J_{EI}$), respectively. The areas marked with Roman numerals represent the three different dynamic modes described above. Specifically, in area (I) the only attractor is a limit cycle, in area (II) there is bistability with a stable limit cycle and a stable state of rest, and in area (III) the only attractor is a state of rest. We see that all modes occupy rather large areas in the parameter spaces, i.e., they are robust to parameter changes in wide intervals. Vertical and horizontal dash-dotted lines show cross-sections of two-parameter bifurcation diagrams, which correspond to one-parameter diagrams presented in Fig.~\ref{bif_onepar3} (see figure caption for details). The intersection of the horizontal and vertical lines represents the parameter values used in Fig.~\ref{free_netw_dyn}.

\section{Suppressing synchronous spiking}
\label{sec:suppression}

\subsection{High-frequency stimulation of the inhibitory population }
\label{sec:inhib_popul}

We first show that synchronous spiking of the network can be effectively suppressed by high-frequency stimulation of the inhibitory population. We consider the network dynamics for $I_E(t)=0$ and 
\begin{equation}
I_I(t)= a \cos(\omega t),  \label{I_I}
\end{equation}
where $a$ is the amplitude and $\omega$ is the angular frequency of HF stimulation. The stimulation current Eq.~\eqref{I_I} satisfies the clinically mandatory charge balance condition $\int_0^T I_I(t)dt =0$, where $T=2\pi/\omega$ is the stimulation period. We assume that the values of the parameters are chosen such that the free network has a single stable attractor of the limit cycle.  We also assume that the stimulation frequency $\nu=1/T$ is considerably greater than the frequency $\nu_0$ of the limit cycle.

\begin{figure}
\centering
	\includegraphics{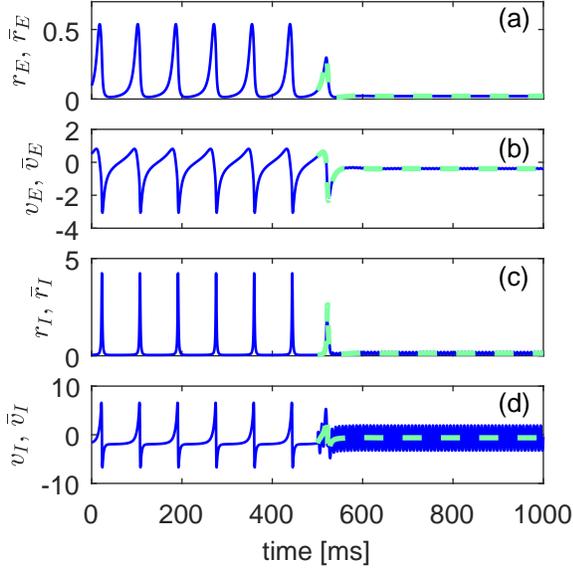}
\caption{\label{netw_dyn_aver_sin} Suppression of network oscillations by HF stimulation of the inhibitory population. Thin solid blue curves show the  dynamics of the  variables $r_{E}(t)$, $v_{E}(t)$, $r_{I}(t)$ and $v_{I}(t)$ obtained by solving the Eqs.~\eqref{eq_rvEI} for the same parameter values as in Fig.~\ref{free_netw_dyn}. For $t<500$ ms, the network is not stimulated and demonstrates exactly the same dynamics as in Fig.~\ref{free_netw_dyn}, and at $t\geq 500$ ms, the HF stimulation of inhibitory neurons is activated with the frequency  $\nu=130$ Hz and the amplitude $a=30$. The bold dashed green curves show dynamics of the variables $\bar{r}_{E}(t)$, $\bar{v}_{E}(t)$, $\bar{r}_{I}(t)$ and $\bar{v}_{I}(t)$ obtained by solving the averaged  Eqs.~\eqref{eq_rvEI_aver}.
}
\end{figure}

A numerical example of the effect of HF stimulation on network dynamics is shown in Fig.~\ref{netw_dyn_aver_sin}. The solid blue curves show the solution of the Eqs~\eqref{eq_rvEI} for the same parameter values as in Fig.~\ref{free_netw_dyn}. For time $t<500$ ms, the network is not perturbed and demonstrates exactly the same dynamics as in Fig.~\ref{free_netw_dyn}: It oscillates at a frequency of $\nu_0=1/T_0 \approx 11.5$ Hz. For $t\geq 500$ ms, HF stimulation of inhibitory neurons is activated with a frequency $\nu=130$ Hz and an amplitude $a=30$. We see that HF stimulation effectively suppresses  synchronized spiking in the network. The spiking rates of excitatory and inhibitory  neurons $r_E$ and $r_I$ drop to almost zero. The mean membrane potential $v_E$ of the excitatory population becomes almost constant. The mean membrane potential $v_I$ of the stimulated inhibitory population shows high-frequency oscillations around the resting state with a reduced amplitude.

To understand why HF stimulation of inhibitory neurons is so effective at suppressing network oscillations, we refer to the method of averaging \cite{sanders07}, which is widely used in various fields of physics including vibrational mechanics. This method makes it possible to explain the effect of oscillation suppression in terms of stabilizing the unstable  resting state of the network. The effect is similar to stabilization of the upside-down position of a rigid pendulum by vibrating its pivot up and down at a suitably high frequency~\cite{Kapitza1951,Butikov2001}. A theoretical approach to solving the pendulum problem was first proposed by Kapitza~\cite{Kapitza1951}. It is based on dividing the dynamics of a pendulum into fast and slow motion and deriving an averaged equation for the slow dynamics of a pendulum. This approach has recently been applied to a single spiking neuron stimulated by a HF field~\cite{Pyragas2013}. Here, we adapt this approach to the network Eqs.~\eqref{eq_rvEI}. 

To apply the averaging method to the Eqs.~\eqref{eq_rvEI}, we rewrite them in a more convenient form.  We denote the dynamic variables of the first three equations, which do not contain the HF stimulation term, by a vector 
\begin{equation}
\bold q=(r_E, v_E, r_I)^T. \label{eq_q}
\end{equation}
Then the Eqs.~\eqref{eq_rvEI} can be formally written as
\begin{subequations}
\label{eq_qvI}
\begin{eqnarray}
\tau\dot{\bold q} & = & \bold G(\bold q, v_I), \label{q_qvI1}\\
\tau\dot{v}_I & = & f(\bold q, v_I)+a\cos(\omega t), \label{eq_vI2}
\end{eqnarray}
\end{subequations}
where $\bold G(\bold q, v_I)$ is a three dimensional vector function defined by the RHS of the first three Eqs.~\eqref{eq_rE}, \eqref{eq_vE} and \eqref{eq_rI} at $I_E(t)=0$ and 
\begin{equation}
f(\bold q, v_I)=\bar{\eta}_I +v_I^2-\pi^2 r_I^2+J_{EI}r_E-J_{II}r_I \label{eq_g}
\end{equation}
is a scalar function defined by the RHS of the last Eq.~\eqref{eq_vI}.

Our aim is to simplify the nonautonomous system Eqs.~\eqref{eq_qvI} for large frequencies $\omega$. Using the small parameter $\varepsilon =(\omega \tau)^{-1} \ll 1$, we seek to eliminate the HF term $a\cos(\omega t)$ and obtain an autonomous system, the solutions of which approximate the original system. First, we change the variables of the system Egs.~\eqref{eq_qvI}:
\begin{subequations}
\label{transform}
\begin{eqnarray}
\bold q(t) &=& \bold Q(t), \label{transform_a} \\
v_I(t) &=&V_I(t)+A\sin(\omega t) \label{transform_b}
\end{eqnarray}
\end{subequations}
with
\begin{equation}
A=a/\omega\tau. \label{Aa}
\end{equation}
As in Ref~\cite{Pyragas2013}, we assume that $A$ is of order $O(1)$ with respect to the perturbation parameter $\varepsilon$. This means that we are considering the case of HF stimulation with large amplitudes, $a \sim O(\varepsilon^{-1})$. Only large amplitudes can have a noticeable effect when stimulated with frequencies significantly higher than the natural frequency of the network. The mathematical justification for this assumption can be found in the Appendix of Ref.~\cite{Ratas2012}. Substituting Eqs.~\eqref{transform} into Eqs.~\eqref{eq_qvI}, we derive the following equations for the new variables $\bold Q(t)$ and $V_I(t)$:
\begin{subequations}
\label{eq_transf}
\begin{eqnarray}
\tau\dot{\bold Q} & = & \bold G(\bold Q, V_I+A\sin(\omega t)), \label{eq_transfa}\\
\tau\dot{V}_I & = & f(\bold Q, V_I+A\sin(\omega t)). \label{eq_transfb}
\end{eqnarray}
\end{subequations}
By rescaling the time variable $\Theta=\omega t$ (here $\Theta$ is the ``fast'' time)  system \eqref{eq_transf} can be transformed to the standard form of equations as typically used by the method of averaging \cite{sanders07}:
\begin{subequations}
\label{eq_transf1}
\begin{eqnarray}
\frac{d\bold Q}{d\Theta} & = & \varepsilon \bold G(\bold Q, V_I+A\sin(\Theta)), \label{eq_transf1a}\\
\frac{d V_I}{d\Theta} & = & \varepsilon f(\bold Q, V_I+A\sin(\Theta)). \label{eq_transf1b}
\end{eqnarray}
\end{subequations}
Due to the small factor $\varepsilon$ on the RHS of the Eqs.~\eqref{eq_transf1}, the variables $\bold Q$ and $V_I$ vary slowly, while the periodic functions in the RHS oscillate fast. In accordance with the averaging method \cite{sanders07}, an approximate solution of the system  Eqs.~\eqref{eq_transf1} can be obtained by averaging the RHS of the system over fast oscillations. Specifically, let us denote the variables of the averaged system as $\bar{\bold q}=(\bar{r}_E, \bar{v}_E, \bar{r}_I)^T$ and $\bar{v}_I$. They satisfy the equations:
\begin{subequations}
\label{eq_averag}
\begin{eqnarray}
\frac{d\bar{\bold q}}{d\Theta} & = & \varepsilon \left\langle\bold G(\bar{\bold q}, \bar{v}_I+A\sin(\Theta))\right\rangle_{\Theta}, \label{eq_averaga}\\
\frac{d \bar{v}_I}{d\Theta} & = & \varepsilon \left\langle f(\bar{\bold q}, \bar{v}_I+A\sin(\Theta))\right\rangle_{\Theta}. \label{eq_averagb}
\end{eqnarray}
\end{subequations}
Here, the angle brackets denote the averaging over the period of the fast time $\langle (\cdots) \rangle_{\Theta} = (1/2\pi)\int_0^{2\pi} (\cdots) d\Theta$. The method of averaging states that the averaged system \eqref{eq_averag} approximates the solutions of the system \eqref{eq_transf1} with the accuracy of $O(\varepsilon)$, i.e., $\bold Q=\bar{\bold q}+O(\varepsilon)$ and $V_I =\bar{v}_I+O(\varepsilon)$. After returning to the original time scale, the averaged system \eqref{eq_averag} takes the following form:
\begin{subequations}
\label{eq_averag1}
\begin{eqnarray}
\tau\dot{\bar{\bold q}}(t) & = &  \left\langle\bold G(\bar{\bold q}(t), \bar{v}_I(t)+A\sin(\Theta))\right\rangle_{\Theta}, \label{eq_averag1a}\\
\tau\dot{\bar{v}}_I(t) & = &  \left\langle f(\bar{\bold q}(t), \bar{v}_I(t)+A\sin(\Theta))\right\rangle_{\Theta}. \label{eq_averag1b}
\end{eqnarray}
\end{subequations}
Here, the dot means differentiation by the original time $t$.
Finally, the solution of the original non-autonomous system \eqref{eq_qvI} can be expressed in terms of the solution of the averaged (autonomous) system \eqref{eq_averag1} as follows:
\begin{subequations}
\label{sol_ap}
\begin{eqnarray}
\bold q(t) &=& \bar{\bold q}(t)+O(\varepsilon),  \label{sol_apa}\\
v_I(t) &=& \bar{v}_I(t)+A\sin(\omega t)+O(\varepsilon).  \label{sol_apb}
\end{eqnarray}
\end{subequations}
The substitution \eqref{transform} and subsequent application of the averaging method allowed us to separate the slow and fast motion of the network and   present the solution in the form of their superposition. The terms $\bar{\bold q}(t)$ and $\bar{v}_I(t)$ in the Eqs.~\eqref{sol_ap} represent slow motion and satisfy the averaged Eqs.~\eqref{eq_averag1} while the term $A\sin(\omega t)$ describes high-frequency oscillations of the mean membrane potential of the stimulated inhibitory population.

After performing the averaging procedure in the Eqs.~\eqref{eq_averag1}, we write them explicitly:
\begin{subequations}
\label{eq_rvEI_aver}
\begin{eqnarray}
\tau\dot{\bar{r}}_E & = & \Delta_E/\pi+ 2\bar{r}_E\bar{v}_E, \label{eq_rvEI_avera}\\
\tau\dot{\bar{v}}_E & = & \bar{\eta}_E +\bar{v}_E^2-\pi^2 \bar{r}_E^2-J_{IE}\bar{r}_I,\label{eq_rvEI_averb}\\
\tau\dot{\bar{r}}_I & = & \Delta_I/\pi+ 2\bar{r}_I\bar{v}_I, \label{eq_rvEI_averc}\\
\tau\dot{\bar{v}}_I & = & \bar{\eta}_I^A +\bar{v}_I^2-\pi^2 \bar{r}_I^2+J_{EI}\bar{r}_E-J_{II}\bar{r}_I. \label{eq_rvEI_averd}
\end{eqnarray}
\end{subequations}
Formally, these equations are similar to the original Eqs.~\eqref{eq_rvEI},
but the HF term $I_I(t)=a\cos(\omega t)$ is excluded from the Eq.~\eqref{eq_vI} (recall that here we are considering the case $I_E(t)=0$).
The only difference between the Eqs~\eqref{eq_rvEI} without stimulation and the averaged Eqs.~\eqref{eq_rvEI_aver} is that the parameter $\bar{\eta}_I$ is replaced by the parameter $\bar{\eta}_I^A$, whose value depends on the stimulation parameter $A$:
\begin{equation}
\bar{\eta}_I^A=\bar{\eta}_I+A^2/2. \label{eq_mdif_eta}
\end{equation}
In Fig.~\ref{netw_dyn_aver_sin}, we compare the solution of the averaged Eqs.~\eqref{eq_rvEI_aver} with the solution of the original Eqs.~\eqref{eq_rvEI}. For $t<500$~ms, there is no stimulation ($a=0$ and $A=0$) and therefore $\bar{\eta}_I^A=\bar{\eta}_I$. In this case, the Eqs.~\eqref{eq_rvEI_aver} and~\eqref{eq_rvEI} are identical and give exactly the same solution. For  $t\ge 500$~ms, when stimulation is activated, the averaged Eqs.~\eqref{eq_rvEI_aver} (bold dashed green curves) also approximate well the solution of the original system Eqs.~\eqref{eq_rvEI} (thin solid blue curves). 

The connection of the averaged mean-field Eqs.~\eqref{eq_rvEI_aver} with the original unperturbed mean-field Eqs.~\eqref{eq_rvEI} allows us to predict the effect of HF stimulation by analyzing the solutions of the unperturbed system and simply explain the oscillation suppression mechanism. According to  Eq.~\eqref{eq_mdif_eta}, the effect of HF stimulation on the averaged dynamics of the network is a change in the parameter $\bar{\eta}_I$, which determines the center of the Lorentzian distribution $g_I(\eta)$ of inhibitory neurons. This center shifts to the right by a distance of $A^2/2$. As a result, the proportion of spiking neurons in the population increases, and the proportion of quenched neurons decreases. Thus, the inhibitory population becomes more active. A sufficient increase in the parameter $\bar{\eta}_I$ can lead to stabilization of the initially unstable resting state of the network and, as a consequence, to the termination of oscillations. 

\begin{figure}
\centering
	\includegraphics{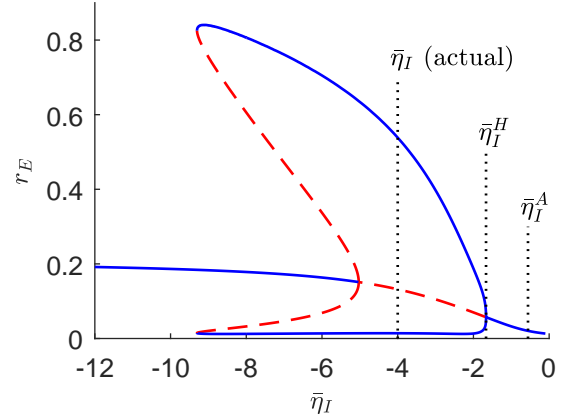}
\caption{\label{bif_etaI_rE} One-parameter bifurcation diagram of the unperturbed network showing the dependence of the firing rate $r_E$ on the parameter $\bar{\eta}_I$. The rest of the parameters are fixed in the same way as in Fig.~\ref{free_netw_dyn}. The designations for the solid blue and dashed red curves are the same as in Fig.~\ref{bif_twopar2}. Three vertical dotted lines show the actual value $\bar{\eta}_I=-4$ of the bifurcation parameter, the value $\bar{\eta}_I^H\approx -1.667$ of the supercritical Hopf bifurcation, and the value $\bar{\eta}_I^A\approx -0.559$ obtained from the Eq.~\eqref{eq_mdif_eta} at stimulation frequency $\nu=130$~Hz and amplitude $a=30$.
}
\end{figure}

The mechanism of stabilization of the resting state is evident from the one-parameter bifurcation diagram $r_E$ versus $\bar{\eta}_I$ of the unperturbed system shown in Fig.~\ref{bif_etaI_rE}. Three vertical dotted lines show the actual value $\bar{\eta}_I=-4$ of the bifurcation parameter, the value $\bar{\eta}_I^H\approx -1.667$ of the supercritical Hopf bifurcation, and the value $\bar{\eta}_I^A\approx -0.559$ obtained from  Eq.~\eqref{eq_mdif_eta} at stimulation frequency $\nu=130$~Hz and amplitude $a=30$. We see that the actual value of $\bar{\eta}_I$ is in the region where the limit cycle is stable and the resting state is unstable. Due to HF stimulation, this value is shifted beyond the Hopf bifurcation point $\bar{\eta}_I^H$ to the position  $\bar{\eta}_I^A$, where the resting state is stable. 

\begin{figure}
\centering
	\includegraphics{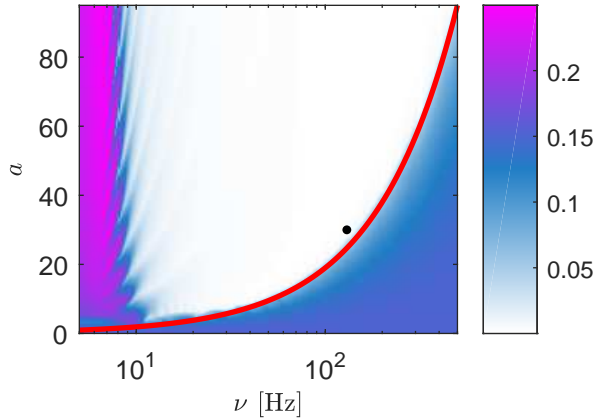}
\caption{\label{a_nu_sig} Network response to stimulation of the inhibitory population depending on the stimulation frequency $\nu$ and the amplitude $a$. The colors show the values of the standard deviation $\sigma$, estimated by the Eq.~\eqref{eq_variance} using a $5000$ ms time window for averaging. White represents an area of suppressed oscillations. Without stimulation ($a=0$), the standard deviation is $\sigma \approx 0.15$.  The solid red curve shows the analytical threshold amplitude Eq.~\eqref{eq_stab_cond}. The black dot denotes the values of the  $(\nu, a)$ parameters used in Fig.~\ref{netw_dyn_aver_sin}.
}
\end{figure}

In the general case, the condition for stabilization of the resting state is $\bar{\eta}_I^A>\bar{\eta}_I^H$ or $A^2>2(\bar{\eta}_I^H-\bar{\eta}_I)$.
Taking into account the Eq.~\eqref{Aa}, this condition can be written as
\begin{equation}
a> a_{\text{th}} \equiv 2\pi \nu \tau \sqrt{2(\bar{\eta}_I^H-\bar{\eta}_I)}, \label{eq_stab_cond}
\end{equation}
where $a_{\text{th}}$ is a threshold amplitude of the HF stimulation. When this amplitude is exceeded, the resting state of the averaged Eqs.~\eqref{eq_rvEI_aver} is stabilized, and the limit cycle oscillations in the original HF stimulated system Eqs.~\eqref{eq_rvEI} are suppressed. Equation~\eqref{eq_stab_cond} shows that the threshold amplitude $a_{\text{th}}$ is proportional to the stimulation frequency $\nu$. Thus, with an increase in the frequency of stimulation, it is necessary to proportionally increase the amplitude of stimulation if we want to maintain a stable state of rest. 
Note that the Eq.~\eqref{eq_stab_cond} is only valid for sufficiently large frequencies $\nu \gg 1/2\pi\tau$. Information about the response of the network to stimulation at low frequencies, which are less or comparable to the limit cycle frequency, can be found by direct integration of the Eqs.~\eqref{eq_rvEI}. Figure~\ref{a_nu_sig} shows the result obtained by integrating the Eqs.~\eqref{eq_rvEI} while changing both stimulation parameters, amplitude $a$ and frequency $\nu$. As a measure of the network's response to stimulation, we choose the standard deviation of the spiking rate of the excitatory population
\begin{equation}
\sigma= \sqrt{\langle\left[r_E(t)-\langle r_E(t)\rangle\right]^2 \rangle}, \label{eq_variance}
\end{equation}
where angle brackets denote time average. Small values of this parameter correspond to a stable resting state of the network, and large values indicate oscillations of large amplitude. In Fig.~\ref{a_nu_sig}, the values of $\sigma$ in the parameter plane  $(\nu, a)$  are shown in colors. White represents the area of suppressed oscillations.  For high frequencies, the boundary of this area is in good agreement with the analytical curve of the threshold amplitude Eq.~\eqref{eq_stab_cond}, which is shown by the solid red curve. For low frequencies, stimulation increases the oscillation amplitude. 
In the region $\nu<8$~Hz, the standard deviation is  significantly larger than the value of $\sigma \approx 0.15$ observed in the network without stimulation. Interestingly, the white area in Fig.~\ref{a_nu_sig} resembles an experimentally obtained area on the plane of the parameters of frequency and intensity of stimulation, where tremor in patients with Parkinson's disease was eliminated by stimulation of the ventral intermediate thalamic nucleus~\cite{Benabid1991}.

\subsection{Suppression of oscillations by controlling the excitatory population }
\label{sec:excit_popul}
 
\begin{figure}
\centering
	\includegraphics{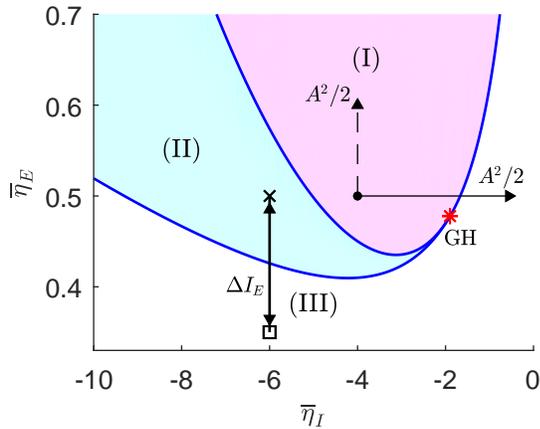}
\caption{\label{Bif_etaI_etaE} Two-parameter bifurcation diagram of the network without stimulation in the plane $(\bar{\eta}_I, \bar{\eta}_E)$. As in Fig.~\ref{bif_twopar2}, the areas marked with Roman numerals correspond to different dynamic modes of the network: (I) -- the only stable limit cycle, (II) -- bistability with a stable limit cycle and a stable state of rest, and (III) -- the only stable state of rest. The red asterisk marked with letters GH denotes the point of the generalized Hopf bifurcation. The black dot in area (I) indicates the values of the parameters used in Fig.~\ref{netw_dyn_aver_sin}. HF stimulating effect of the inhibitory population is shown by the solid horizontal arrow. The vertical dashed arrow shows the effect of HF stimulation of the excitatory population. The cross in area (II) indicates the values of the parameters used in Fig.~\ref{pulse_contr}. The double vertical arrow connecting the cross to the square corresponds to the inhibitory pulse applied to the excitatory population. The pulse dynamics is shown in Fig.~\ref{pulse_contr}(e). 
}
\end{figure}

Next, we will consider the possibility of suppressing network oscillations by HF stimulation of the excitatory population. Now we put $I_I(t)=0$ and
\begin{equation}
I_E(t)= a \cos(\omega t).  \label{I_E}
\end{equation}
Applying the averaging method to the Eqs.~\eqref{eq_rvEI}, we can derive an autonomous system of averaged equations, similar to the Eqs.~\eqref{eq_rvEI_aver}, but now the parameter $\bar{\eta}_I$ remains unchanged, and the parameter $\bar{\eta}_E$ will be modified as
\begin{equation}
\bar{\eta}_E \rightarrow \bar{\eta}_E^A=\bar{\eta}_E+A^2/2. \label{eq_mdif_etaE}
\end{equation}
An increase in the $\bar{\eta}_E$ parameter means that  the proportion of spiking neurons in the excitatory population increases, while the proportion of quenched neurons decreases. As a result, the excitatory population becomes more active and, in contrast to the case of stimulation of the inhibitory population, the oscillatory effects are now enhanced. Figure~\ref{Bif_etaI_etaE} provides a graphical explanation of why the effects of HF stimulation of excitatory and inhibitory populations are different. Here, we show a two-parameter bifurcation diagram of the network without stimulation in the plane $(\bar{\eta}_I, \bar{\eta}_E)$. As in Fig.~\ref{bif_twopar2}, the areas marked with Roman numerals correspond to different dynamic modes of the network: (I) -- the only stable limit cycle, (II) -- bistability  and (III) -- the only stable state of rest. The black dot in area (I) indicates the actual values of the parameters. The stimulating effect of the inhibitory population is shown by the solid horizontal arrow. As a result of stimulation, the dot showing the actual values of the parameters is shifted to the right to the region (III), where the state of rest is stable and, thus, the oscillations are suppressed. The stimulating effect of the excitatory population is shown by the dashed vertical arrow. Now the dot shifts upward and remains in area (I), where the limit cycle is the only attractor and, thus, the oscillations are preserved.
\begin{figure}
\centering
	\includegraphics{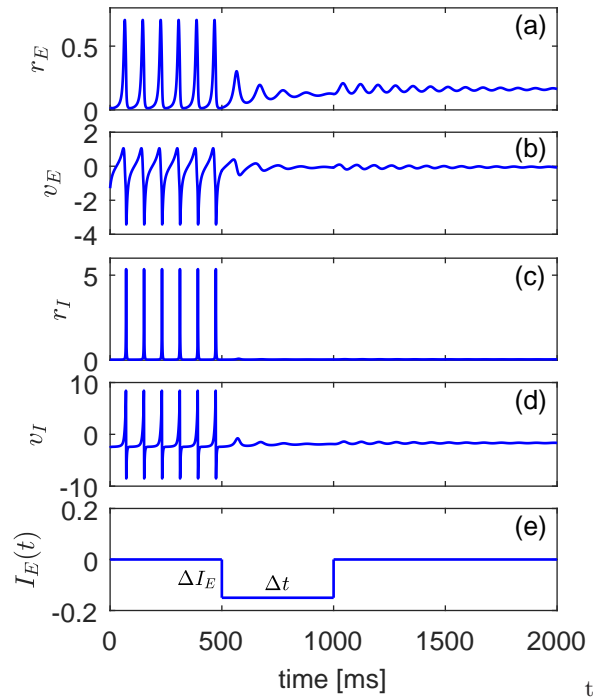}{t}
\caption{\label{pulse_contr} Elimination of network oscillations using an inhibitory pulse applied to the excitatory population. Dynamics of  variables (a) $r_{E}(t)$, (b) $v_{E}(t)$, (c) $r_{I}(t)$ and (d) $v_{I}(t)$ obtained by solving the Eqs.~\eqref{eq_rvEI} with the inhibitory pulse $I_E(t)$ shown in (e). Pulse amplitude is $\Delta I_E=-0.15$ and duration $\Delta t=500$ ms. The parameters except for $\bar{\eta}_I=-6$ are the same as in Fig.~\ref{free_netw_dyn}. 
}
\end{figure}

Although HF stimulation of the excitatory population is ineffective in suppressing network oscillations, in some cases we can still eliminate oscillations by applying a different type of control signal to the excitatory population. This is possible if the network parameters are in the bistable area, and the network is in oscillatory mode. In this case, the network can be switched from stable periodic oscillations to a stable state of rest by applying a single inhibitory rectangular pulse to the excitatory population. The idea of this control algorithm is graphically illustrated in Fig.~\ref{Bif_etaI_etaE}. We assume that the actual values of the parameters $(\bar{\eta}_I, \bar{\eta}_E)$ without stimulation are located in the bistable area (II). They are marked with a cross. We suppose that the network is in the oscillatory mode. We then apply a negative rectangular pulse $I_E(t)$ with amplitude $\Delta I_E <0$ and duration $\Delta t$ to the excitatory population, as shown in Fig.~\ref{pulse_contr} (e). Since the parameters $I_E$ and $\bar{\eta}_E$ enter into  the Eq.~\eqref{eq_vE} as a sum, we can interpret this pulse as it applies to the $\bar{\eta}_E$  parameter not to  $I_E$. In Fig.~\ref{Bif_etaI_etaE}, we demonstrate this interpretation using the vertical double arrow connecting the cross with the square, which shows the values of the $(\bar{\eta}_I, \bar{\eta}_E)$ parameters when $I_E=\Delta I_E$. For sufficiently large pulse amplitude $\Delta I_E$, parameter values marked with a square appear in area (III), where the fixed point is the only attractor. If the pulse duration $\Delta t$ is long enough, the system will approach the fixed point. We expect that for two different values of the parameters marked with a cross ($I_E = 0$) and a square ($I_E = \Delta I_E$), the coordinates of fixed points in the phase space are close to each other, so that at the end of the pulse the state of the system will be in the basin of attraction of a fixed point corresponding to the value $I_E = 0$.  Then, being in the bistable area, the system will approach a stable fixed point and remain at rest at zero stimulation current $I_E = 0$.

In Fig.~\ref{pulse_contr}, we demonstrate the efficiency of this algorithm for $\bar{\eta}_I=-6$ and other parameters such as in Fig.~\ref{free_netw_dyn}. This set of parameters is in the bistable area (II), as shown in Fig.~\ref{Bif_etaI_etaE}. Figures~\ref{pulse_contr}(a), \ref{pulse_contr}(b), \ref{pulse_contr}(c) and \ref{pulse_contr}(d) show, respectively, the dynamics of variables $r_E$, $v_E$, $r_I$ and $v_I$ obtained by integrating the Eqs.~\eqref{eq_rvEI} with inhibitory pulse $I_E(t)$ shown in Fig.~\ref{pulse_contr}(e). As expected, this pulse stops the oscillation in the network, and the network remains at rest without further stimulation. Note that unlike the HF current Eq.~\eqref{I_I}, the inhibitory pulse used here does not satisfy the charge balance condition. However, this condition is not required for single pulse stimulation.

\section{Modeling microscopic dynamics}
\label{sec:microscopic_model}

The reduced mean-field Eqs.~\eqref{eq_rvEI} are derived in the limit of an infinite-size network, while realistic networks consist of a finite number of neurons. To test if the control algorithms described above work for networks of finite size, here we perform a direct numerical simulation of the microscopic dynamics described by the Eqs.~\eqref{model}. 

Numerical simulation of the Eqs.~\eqref{model} is more convenient after changing the variables
\begin{equation}
V_j^{(E,I)} = \tan(\theta_j^{(E,I)}/2)\label{eq_transf_tet}
\end{equation}
that turn QIF neurons into theta neurons. Such a transformation of variables avoids the problem associated with jumps of infinite size (from $+ \infty $ to $-\infty $) of the membrane potential $V_j^{(E,I)}$ of the QIF neuron at the moments of firing. The phase $\theta_j^{(E,I)}$ of the theta neuron simply crosses the value of $\theta_j^{(E,I)}=\pi$ at these moments. For theta neurons, the Eqs.~\eqref{model} are transformed into
%
%\begin{equation}
\begin{eqnarray}
\tau \dot{\theta}_{j}^{(E,I)}&=& 1-\cos \left(\theta_{j}^{(E,I)}\right)\nonumber\\
&+&\left[1+\cos \left(\theta_{j}^{(E,I)}\right)\right]\left[\eta_{j}^{(E,I)}+\mathcal{I}_{j}^{(E,I)} \right]. \label{theta_j}
\end{eqnarray}
%\end{equation}
%
These equations were integrated by the Euler method with a time step of $d t = 5\times 10^{-4}$. Two populations of theta excitatory and inhibitory neurons each consisting of $N=2000$ units with the Lorentzian distributions \eqref{Lor} were deterministically generated using $\eta_j^{(E,I)}=\bar{\eta}_{E,I}+\Delta_{E,I} \tan\left[(\pi/2)(2j-N-1)/(N+1)\right]$,   $j=1,\ldots, N$. More information on numerical modeling of Eqs.~\eqref{theta_j} can be found in Ref.~\cite{Ratas2016}. To compare the results obtained from the microscopic model Eqs.~\eqref{theta_j} with the solutions of the reduced system Eqs.~\eqref{eq_rvEI}, we calculate the Kuramoto order parameters~\cite{Kuramoto2003}
\begin{equation}
\label{eq_Z}
Z_{E,I}=\frac{1}{N}\sum\limits_{j=1}^{N}\exp(i \theta_j^{E,I})
\end{equation} 
for each population and use the relationship between $Z_{E,I}$ and the spiking rate $r_{E,I}$~\cite{Montbrio2015}: 
\begin{equation}
\label{eq_W}
r_{E,I}=\frac{1}{\pi}\operatorname{Re}\left(\frac{1-Z_{E,I}^*}{1+Z_{E,I}^*}\right),
\end{equation} 
where $Z_{E,I}^*$ means complex conjugate of $Z_{E,I}$. 

\begin{figure}
\centering
	\includegraphics{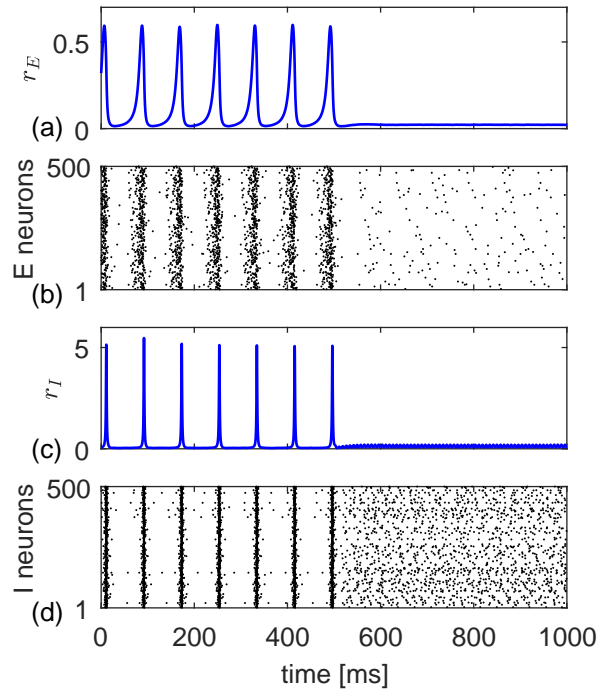}
\caption{\label{netw_dyn_micro_sin} The effect of HF stimulation of the inhibitory population, obtained using the microscopic model Eqs.~\eqref{theta_j}. The number of neurons in the excitatory and inhibitory populations is the same, $N = 2000$. All parameters are the same as in Fig.~\ref{netw_dyn_aver_sin}, and HF stimulation with the amplitude $a=30$ and the frequency $\nu=130$ Hz is also turned on at $t=500$ ms, as in Fig.~\ref{netw_dyn_aver_sin}. (a) and (c) Dynamics of the spiking rates of the populations E and I, respectively. (b) and (d) Raster plots of $500$ randomly selected neurons in populations E and I, respectively. Here, the dots show the spike moments for each neuron, where the vertical axis indicates neuron numbers.
}
\end{figure}

In Fig.~\ref{netw_dyn_micro_sin}, we show the results of HF stimulation of the inhibitory population, obtained from the Eqs.~\eqref{theta_j}. The parameter values are the same as in Fig.~\ref{netw_dyn_aver_sin}. Here, as in Fig.~\ref{netw_dyn_aver_sin}, HF stimulation with an amplitude $a=30$ and a frequency $\nu=130$ Hz is activated at $t>500$ ms. We see that the dynamics of the spiking rates of excitatory and inhibitory populations shown in Figs.~\ref{netw_dyn_micro_sin}(a) and \ref{netw_dyn_micro_sin}(c) are similar to those shown in Figs.~\ref{netw_dyn_aver_sin}(a) and~\ref{netw_dyn_aver_sin}(c), respectively. Thus, the mean-field Eqs.~\eqref{eq_rvEI} are robust, they predict well the macroscopic dynamics of a finite size network consisting of $N=2000$ neurons in each population. Microscopic network behavior can be seen in Figs.~\ref{netw_dyn_micro_sin}(b) and \ref{netw_dyn_micro_sin}(d), which show raster plots of $500$ randomly selected neurons in populations E and I, respectively. Without stimulation ($t<500$ ms),  most neurons in excitatory and inhibitory populations exhibit coherent behavior and produce macroscopic periodic oscillations. HF stimulation ($t>500$ ms) increases the number of active neurons in the inhibitory population, which destroys the coherent spiking in both populations, and the initially unstable incoherent resting state is stabilized. 

\begin{figure}
\centering
	\includegraphics{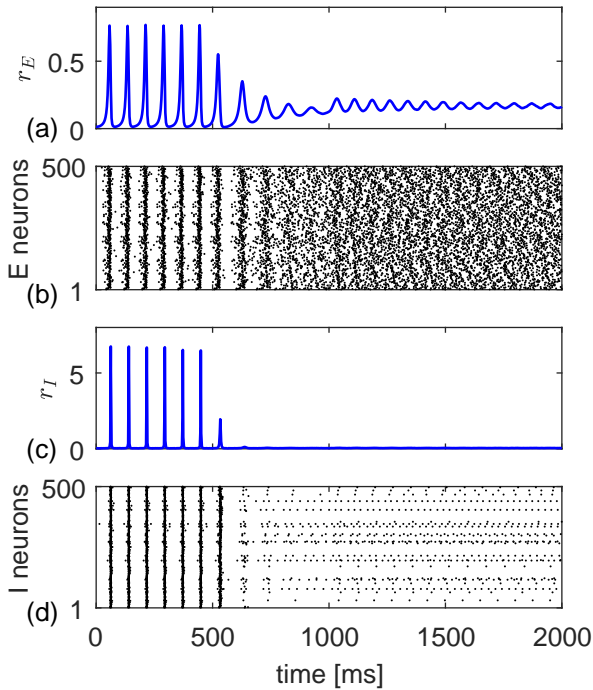}
\caption{\label{netw_dyn_micro_imp} The effect of an inhibitory pulse applied to the excitatory population, obtained from the microscopic model Eqs.~\eqref{theta_j}. The number of neurons in the excitatory and inhibitory populations is $N = 2000$. All parameters are the same as in Fig.~\ref{pulse_contr} and the pulse shape is the same as in Fig.~\ref{pulse_contr}(e). (a) and (c) Dynamics of the spiking rates of the populations E and I, respectively. (b) and (d) Raster plots of $500$ randomly selected neurons in populations E and I, respectively. 
}
\end{figure}

Figure~\ref{netw_dyn_micro_imp} shows the effect of an inhibitory pulse applied to the excitatory population, obtained from the microscopic model Eqs.~\eqref{theta_j}. The values of the parameters are the same as in Fig.~\ref{pulse_contr} and the pulse shape is the same as in Fig.~\ref{pulse_contr}(e). Again we see that the mean-field Eqs.~\eqref{eq_rvEI} predict well the macroscopic dynamics of a finite size network described by Eqs.~\eqref{theta_j}. The dynamics of the spiking rates of excitatory $r_E(t)$ and inhibitory $r_I(t)$ populations shown in Figs.~\ref{netw_dyn_micro_imp}(a) and \ref{netw_dyn_micro_imp}(c) are similar to those presented in Figs.~\ref{pulse_contr}(a) and~\ref{pulse_contr}(c), respectively. Raster plots in Figs.~\ref{netw_dyn_micro_imp}(b) and~\ref{netw_dyn_micro_imp}(d) show the microscopic network dynamics. Before the pulse ($t<500$ ms), most neurons in the excitatory and inhibitory populations behave coherently and produce macroscopic periodic oscillations, which represent one of the two stable modes of the network. A negative pulse with amplitude $\Delta I_E =-0.15$, lasting in the interval 500~ms~$ <t <$~1000~ms, transfers the system to an incoherent resting state, which is the only stable state for the given values of the parameters. For $t>1000$ ms, when the pulse is off, the network remains in a stable incoherent resting state. Note the different point densities on raster plots of populations E and I in incoherent state. This is due to the fact that in population E, a greater proportion of neurons generates spikes than in population I.

\section{Discussion}
\label{sec:conclusions}

We have analyzed the dynamics of a free and stimulated network of two globally connected neural populations consisting of excitatory and  inhibitory quadratic integrate-and-fire neurons. Interaction within and between populations is provided by instantaneous pulses. Both populations are heterogeneous and contain a mixture of at-rest but excitable neurons as well as spontaneously spiking neurons. The heterogeneity is determined by the Lorentzian distribution of the excitability parameter. A model built on these assumptions has two important advantages. First, in the limit of an infinite number of neurons, an exact system of low-dimensional mean-field equations can be obtained. Second, the mean-field equations represent a universal macroscopic model for a large class of neural networks, because they are derived from the microscopic dynamics of QIF neurons, which are the normal form of class I neurons. In contrast to phenomenological neural mass models~\cite{Destexhe2009}, the mean-field equations considered here accurately reproduce the dynamics of spiking neurons for any degree of synchronization and can be considered as next generation neural mass models~\cite{Coombers2019}.

Relatively simple mean field equations make it possible to conduct a thorough bifurcation analysis of various dynamic modes of a free network and to reveal the mechanisms of action of various stimulation algorithms. We performed a bifurcation analysis of a free network depending on the coupling strengths $J_{EI}$ and $J_{IE}$  of the bidirectional interaction between excitatory and inhibitory populations and the coupling strength $J_{II}$, which determines the interaction within the inhibitory population. We also built a bifurcation diagram in the plane of the parameters $(\bar{\eta}_I, \bar{\eta}_E)$, which determine the centers of the distributions of the excitability parameter for inhibitory and excitatory populations. As a result of this analysis, three different modes were established. Depending on the values of the parameters, the system can have a single stable fixed point, a single stable limit cycle, or be in a bistable mode with these two coexisting attractors. All three modes occupy rather large areas in the parameter spaces, which means that they are robust to parameter changes in wide intervals.

As the next step in our analysis, we looked at the problem of controlling   network synchronization. Pathological synchronized oscillations can be the cause of various neurological diseases, and attempts are made to suppress them using external stimulation. Some neurological diseases are successfully treated with high-frequency stimulation. Here, we tested the effectiveness of the HF algorithm for suppressing synchronous spiking in the network of excitatory and inhibitory QIF neurons. We have shown that HF stimulation of the inhibitory population is very effective, whereas HF stimulation of the excitatory population cannot suppress the oscillations. The mechanism of action of HF stimulation is explained using mean-field equations averaged over the stimulation period. The averaged mean-field equations are equivalent to the free mean-field equations, but with a modified  parameter $\bar{\eta}_I$ or $\bar{\eta}_E$, depending on which inhibitory or excitatory population is stimulated.  When HF stimulation is applied to the inhibitory population, changing the $\bar{\eta}_I$ parameter  increases the proportion of spiking neurons in that population. This leads to the stabilization of the state of rest of the network and the termination of oscillations.  The averaged mean-field equations made it possible to obtain an analytical expression for the threshold amplitude of HF stimulation, which stabilizes the resting state. This amplitude is proportional to the frequency of stimulation.
 
HF stimulation of the excitatory population is ineffective, since changing the $\bar{\eta}_E$ parameter  increases the proportion of spiking neurons in the excitatory population and cannot stabilize the resting state of the network. Nevertheless, stopping the network oscillation by controlling the excitatory population can still be achieved if the system parameters are in the bistable area. By applying a rectangular inhibitory pulse to this population, the network state can be switched from the stable limit cycle to the stable state of rest. Such a pulse moves the system parameters to the area where the state of rest is the only attractor, and then returns them back to the bistable area. As a result, the system approaches the stable state of rest, being in the bistable region, and remains in this state in the absence of stimulation. 

To test the performance of the above stimulation algorithms for finite-size networks, we numerically simulated the equations of the microscopic model.
Modeling networks with 2000 excitatory and 2000 inhibitory QIF neurons gave results that are in good agreement with the results obtained from the mean-field equations. Based on our research, we believe that mean-field equations derived from the microscopic dynamics of interacting QIF neurons can serve as an effective tool for developing various stimulation algorithms to control synchronization processes in large-scale neural networks.

\begin{acknowledgements}
This work was supported by Grant No. S-MIP-21-2 of the Research Council of Lithuania.
\end{acknowledgements}

\bibliography{E_I_QIF_network}

\end{document}